\documentclass{WileyChemistry-template}
\usepackage{amsmath}
\usepackage{graphicx}
\usepackage{float}
\usepackage{threeparttable} 
\usepackage{pdfpages}

\title{\raggedright Theoretical exploration of Be--Ag(II)--F phases and their magnetic properties using learning algorithms}

\author{
\begin{minipage}{\textwidth}
	Katarzyna Kuder,\textsuperscript{[a]} Wojciech Grochala,\textsuperscript{[a]} 
\end{minipage}
}

\newcommand{\affiliation}{
\begin{itemize}

\item[{[a]}] K. Kuder, Prof. W. Grochala \\
Centre of New Technologies, University of Warsaw, Banacha 2c St., 02-097 Warsaw, Poland\\
E-mail: k.kuder@cent.uw.edu.pl\\
E-mail: w.grochala@cent.uw.edu.pl
\end{itemize}
}

\renewcommand{\dedication}{
	\begin{minipage}{\textwidth}
		Dedicated to Dr. Zoran Mazej at his 60$^{th}$ birthday
	\end{minipage}
}

\renewcommand{\abstract}{The search for novel silver(II) fluorides is driven by their potential as electronic and magnetic analogues to high-temperature cuprate(II) superconductor precursors. Here, we explore the previously uncharted Be--Ag(II)--F chemical space using global structure prediction algorithms combined with first-principles calculations. Focusing on the AgBeF$_4$ stoichiometry, we identify the five lowest-enthalpy polymorphs crystallizing in the $C$2, $P\bar{1}$, and $P2_1/c$ space groups. All polymorphs show an antiferromagnetic ground state, with AgBeF$_4$\_4 and AgBeF$_4$\_5 exhibiting unprecedented strong superexchange interactions of  $J\approx \text{--460meV}$ and $J\approx \text{--359meV}$ respectively. Those high J values are due to the presence of either [Ag$_2$F$_7$] for AgBeF$_4$\_4, or related infinite $[AgF_{2/2+2/1}]^{2-}$ chains for AgBeF$_4$\_5. Although the phases are found to be metastable with respect to binary difluorides, the thermodynamic analysis suggests that they could be targeted via synthetic routes employing fluorine radicals, with reaction enthalpies reaching $-370\text{ kJ/mol}$.
}

\newcommand{\keywords}{crystal structure prediction \textbullet\ 
	density functional theory \textbullet\ 
	magnetic properties \textbullet\ 
	silver(II) \textbullet\ beryllium
}

\begin{document}

\twocolumn[\vspace{-1.5cm}\maketitle\vspace{-1cm}
	\textit{\dedication}\vspace{0.4cm}]
\small{\begin{shaded}
		\noindent\abstract
	\end{shaded}
}

\begin{figure} [!b]
\begin{minipage}[t]{\columnwidth}{\rule{\columnwidth}{1pt}\footnotesize{\textsf{\affiliation}}}\end{minipage}
\end{figure}




\section*{Introduction}
\label{introduction}

Magnetic materials constitute a central area of modern materials science due to their fundamental importance and wide spectrum of technological applications, such as spintronics, data storage, and quantum information processing \cite{17, 18, 19, 20}. Their physical properties arise from a complex interplay between crystal structure, electronic configuration, and exchange interactions \cite{21}, which makes the prediction and bottom-to-top design of new magnetic materials inherently challenging. In this context, computational approaches have become indispensable, enabling the exploration and screening of novel compounds prior to experimental realization \cite{22, 23, 24}. Usefulness of the modelling approach is particularly large for coordination polymers which do not contain any rigid organic elements but rather are exclusively composed of inorganic moieties.

Among transition-metal-based systems, compounds containing Ag(II) cations are of particular interest due to the $d^9$ electronic configuration of the silver(II) ion, which is isoelectronic with Cu(II). This correspondence is especially significant in the context of cuprate physics, where the $d^9$ configuration plays a central role in the emergence of magnetic phenomena \cite{34, 35, 36, 37}. Importantly, magnetic fluctuations play a crucial role for the emergence of superconductivity in doped cuprates(II).
Ag(II) fluorides - oxocuprates(II) analogs - are especially compelling in this regard, as the high electronegativity of fluorine stabilizes the uncommon high oxidation state of silver while promoting significant metal--ligand covalence \cite{27}. Previous studies have revealed that even subtle structural distortions in silver(II) fluorides can lead to a diverse range of possible new magnetic materials \cite{10,12, 11, 25, 26}.

Although Ag(II) forms ca. 100 distinct fluoride connections with other elements, many ternary systems still constitute an uncharted territory. E.g., only recently we have gained a theoretical insight into Li(I)-Ag(II)-F phase diagram and identified new metastable phases (\ce{LiAgF3},\\ \ce{Li2AgF4}) which could be targetted in experiments \cite{54}. Replacing Li(I) with Be(II) is even more interesting due to a marked covalence of Be(II)-F bonding, which in turn influences the rigidity of crystalline networks. We note that beryllium fluoride shows rich silica-like polymorphism with its highly stable [BeF$_4$] tetrahedral building blocks present in the extended polymeric frameworks \cite{28, 29, 30}. The rigid anionic BeF$_{4/2}$ manifolds could thus act as attractive scaffolds for stabilizing electronically active cationic centers and for engineering mixed-metal fluoride architectures \cite{31, 32}. 

From a crystal-chemical perspective, combining Be--F\\frameworks with Ag(II) units might open a pathway towards phases in which magnetic properties could exist in a  nontrivial way \cite{33}. Getting insight into \ce{BeF2}/\ce{AgF2} chemistry could also be important for understanding of the chemical identity of phases which are responsible for Meissner-effect-like magnetic anomalies, reminiscent of superconductivity, which have been previously observed in the \ce{AgF2}/\ce{BeF2} system \cite{33}.

The Be–Ag–F chemical space has so far remained largely unexplored. In particular, there is no systematic assessment of which Be–Ag–F compositions may be thermodynamically viable and what structural motifs they may adopt, as well as what magnetic and electronic properties they could have.

In this work, we address this gap through an exploration of one particular Be--Ag(II)--F stoichiometry stoichiometry, i.e. Ag(II)BeF$_4$. The search for its possible crystal structures was performed using a global structural prediction approach with~~self-learning algorithms \cite{1, 2, 16}, enabling an efficient exploration of the configurational space and identification of low-enthalpy candidate structures. By combining global structure prediction with the density functional theory methods, this work provides a systematic assessment of the structural, electronic, and magnetic properties of the most stable Ag(II)BeF$_4$ phases within a unified computational framework.

\begin{figure*}[ht]
\begin{center}
\includegraphics[width=17.2cm]{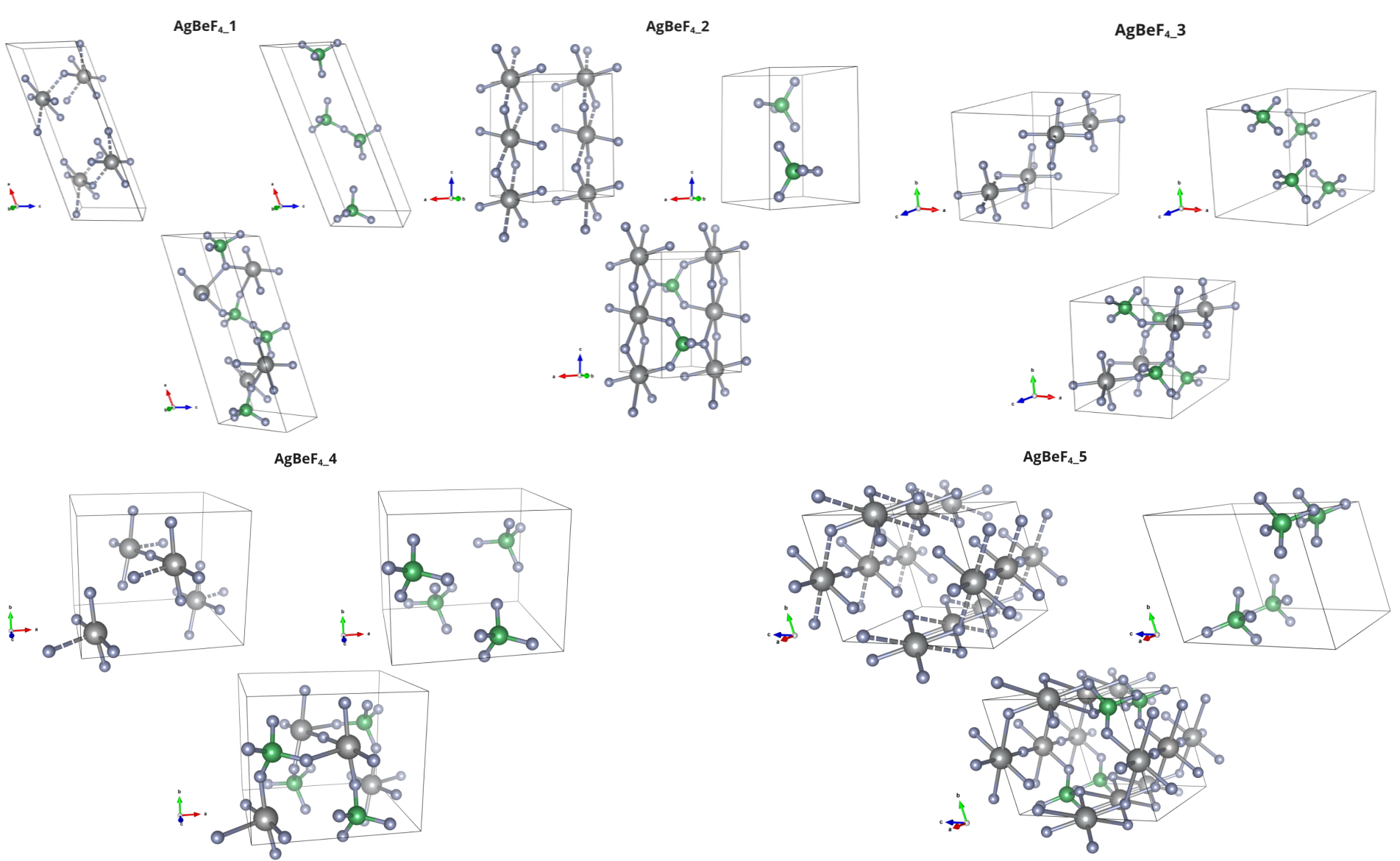}
\caption{Coordination environments within the five proposed AgBeF$_4$ polymorphs. For each structure, the coordination sphere of silver(II) is shown first, followed by the coordination sphere of beryllium, and finally a combined view illustrating both environments. Dark grey spheres represent Ag$^{(2+)}$, light grey spheres correspond to fluorine atoms, and green spheres denote beryllium atoms. Short and long Ag-F bonds are shown using solid and broken lines, respectively.}
\label{F1}
\end{center}
\end{figure*}

\section*{Results and Discussion}
\label{results_discussion}

\subsection*{Crystal structures}

The learning-algorithm quest has produced a few thousand distinct crystal structures of AgBeF$_4$, corresponding to many unique polymorphs. Our attention here will focus on the five lowest-enthalpy structures. As shown in Table \ref{T1}, the lowest-energy structure AgBeF$_4$\_1 as well as AgBeF$_4$\_5 crystallize in the polar \textit{C}2 space group, while the other polymorphs crystallize in the most common centrosymmetric $P\bar{1}$, \textit{P}2$_1$/c space groups. The absence of inversion symmetry in the \textit{C}2 phases suggests the possibility of important functional properties such as piezoelectricity or nonlinear optical response.

\begin{table}[h]
\centering
\caption{Lattice parameters and space groups for AgBeF$_4$ structures.}
\label{T1}
\resizebox{\textwidth}{!}{
\begin{tabular}{c c c c c c c c}
\hline
\textbf{Structure}& 
\begin{tabular}{c}\textbf{Space}\\[-2pt]\textbf{group}\end{tabular}&
\textbf{$a$ (\AA)} &
\textbf{$b$ (\AA)} &
\textbf{$c$ (\AA)} &
\textbf{$\alpha$ ($^{\circ}$)} &
\textbf{$\beta$ ($^{\circ}$)} &
\textbf{$\gamma$ ($^{\circ}$)}\\
\hline
AgBeF$_4\_1$   & \textit{C}2        & 13.522 & 4.920 & 5.085 & 90.0 & 109.6 & 90.0 \\
AgBeF$_4\_2$  & $P\bar{1}$ & 4.815 & 4.825  &  7.205 & 93.7 & 93.6 &  110.7 \\
AgBeF$_4\_3$ & \textit{P}2$_1$/c     & 6.239 &  7.301  & 7.251 & 90.0 & 117.6 & 90.0 \\
AgBeF$_4\_4$ & \textit{C}2     & 7.697 &  6.444  & 7.130 & 90.0 & 90.1 & 90.0 \\
AgBeF$_4\_5$ & $P\bar{1}$   & 4.080 &  6.136  & 7.170 & 75.0 & 81.1 & 85.7 \\
\hline
\end{tabular}
}
\end{table}

Coordination sphere of silver(II) for each explored AgBeF$_4$ structure is very similar and it resembles an elongated octahedron or a distorted tetragonal pyramide (for AgBeF$_4\_4$). Such local coordination of Ag(II) is typical for Ag(II)--F containing systems due to the presence of a strong Jahn--Teller effect  \cite{11,25,38,39}. Similarly, in each proposed structure, as shown in Figure \ref{F1}, beryllium cations adopt a tetrahedral first coordination sphere characteristic of BeF$_2$ \cite{28, 29, 30} and \ce{BeF4}$^{2-}$ salts \cite{41, 42, 43}.

\begin{figure*}[ht]
\begin{center}
\includegraphics[width=13cm]{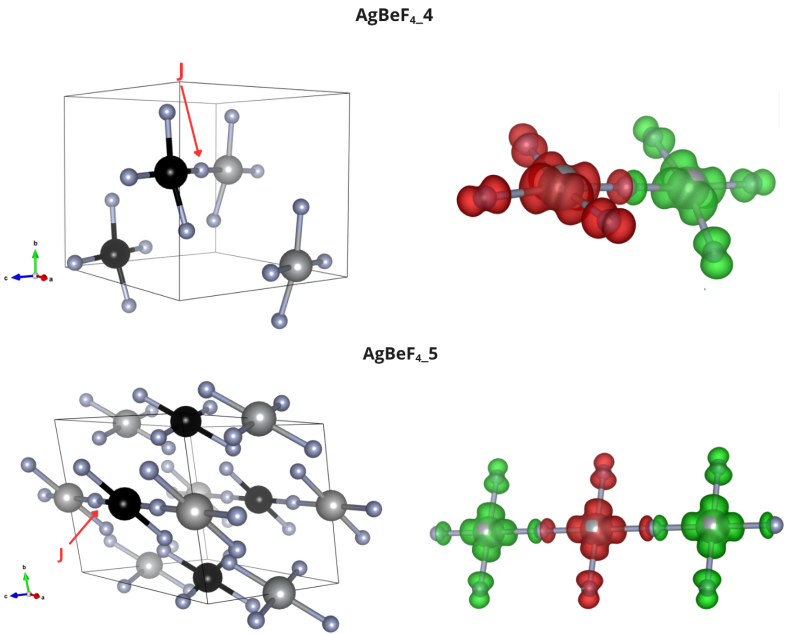}
\caption{The ground state antiferromagnetic models investigated for the AgBeF$_4$\_4 and  AgBeF$_4$\_5 polymorphs.  Spin-up silver(II) cations are represented by grey spheres, while spin-down cations are shown as black spheres. To simplify the pictures, lithium cations and their coordination spheres were omitted. The visualization of the spin density within the infinite chain is also presented with the cutoff value of 0.005e/A$^3$ \cite{53}.}
\label{F3}
\end{center}
\end{figure*}

The silver(II) cation in the lowest enthalpy structure \\adopts a distorted octahedral coordination sphere, with four short Ag$^{(2+)}$--F bonds ranging from 2.031\AA~ to 2.105\AA~, and two longer ones, one of which is significantly longer than the other: 2.513\AA~ and 2.990\AA. As shown in Figure \ref{F1} the octahedra in this structure are irregular. The shortest Ag(II)--F bonds form tilted $[AgF_{2/2+2/1}]^{2-}$ chains (as shown in the Figure \ref{F3}), which are also present in ambient and high-pressure polymorphs of X$_2$AgF$_4$ (where X is an alkali metal) \cite{39}. The \ce{Ag(II)\bond{-}F\bond{-}Ag(II)}
 angle is 135.9$^\circ$.

For the second proposed structure AgBeF$_4$\_2, silver(II)--fluorine bonds range from 2.054\AA~~to 2.099\AA~~for the four short ones, and, two longer of 2.551\AA. In contrast to the first structure, this one exhibits isolated $[AgF_4]^{2-}$ squares within the closest Ag(II)--F bonds. The \ce{Ag(II)\bond{-}F\bond{-}Ag(II)}  angle between one long and one short Ag--F bond is 101.2$^\circ$. Similarly to AgBeF$_4$\_2, AgBeF$_4$\_3 structure also has  four short silver(II)--fluorine bonds which lead to the formation of isolated squares (2.066\AA~~--~~2.096\AA). The two longer bonds are 2.550\AA~~and 2.628\AA. The \ce{Ag(II)\bond{-}F\bond{-}Ag(II)}  angle between one long and one short Ag--F bond is 102.2$^\circ$ (similar to AgBeF$_4$\_2). Such mutual arrangement of the isolated squares is reminiscent of that found for Na2AgF4 \cite{72}.

The AgBeF$_4$\_4 structure, in contrast to the structures described above, features [Ag$_2$F$_7$] dimers. They are formed by the connection of two [AgF$_4$]$^{2-}$ squares, which are slightly tilted. Within the [Ag$_2$F$_7$] dimers, the Ag(II)--F bonds are very short and range from 2.003\AA~~to 2.098\AA. In this polymorph, there is only one long silver(II)--fluorine bond (2.633\AA). The\ce{Ag(II)\bond{-}F\bond{-}Ag(II)} angle in this structure is 178.3$^\circ$. The shortest Ag-F bond distances within the [Ag$_2$F$_7$] dimers resemble those found in the structure of \ce{Ag2ZnZr2F14} (2.017\AA~~to 2.160\AA) \cite{56}, in the HP2 high-pressure structure of \ce{AgF2} (2.084\AA~~to 2.118\AA) \cite{55}, and in the recently proposed structure of LiAgF$_3$\_2 polymorph (2.012\AA~~to 2.084\AA) \cite{54}.

In the structure of the last polymorph, AgBeF$_4$\_5, there are two types of silver(II) cations: one forms the infinite chains, where the four short Ag--F bonds are 2.040\AA~~and 2.068\AA, and the two long bonds are 2.581\AA; whereas the second type forms isolated [AgF$_4$]$^{2-}$ units, in which the short bonds are 2.046\AA~~and 2.070\AA, while the two very long contacts are 2.992\AA. In contrast to the previously discussed structures the \ce{Ag(II)\bond{-}F\bond{-}Ag(II)} angle within the infinite $[AgF_{2/2+2/1}]^{-}$ chain is 180.0$^\circ$. Such perfectly straight chains composed of equiplanar [AgF$_4$] squares are unprecedented in the entire Ag(II) fluoride chemistry, as so far only the nonequiplanar (90$^\circ$-tilted) square geometry was seen for\\ \ce{CsAgF3} \cite{57}. On the other hand, infinite chains seen for \ce{AgFBF4} exhibit linear rather than square planar coordination  of Ag(II) \cite{58}. As we will see, this has enormous consequences for the magnetic and electronic structure of AgBeF$_4$\_5. The $[AgF_{2/2+2/1}]^{-}$ chain is analogous to the $[CuO_{2/2+2/1}]^{4-}$ chain present in the structure of Sr$_2$CuO$_3$ oxocuprate.

The bonds within the BeF$_4^{2-}$ tetrahedra for all polymorphs range from 1.537\AA to 1.595\AA;  their average of 1.566\AA is somewhat larger than those found in the BeF$_2$ structure (1.542-1.547\AA) \cite{40}. Depending on the structure, the BeF$_4^-$ tetrahedra are either isolate, form dimers, or even chains. These entities serve as linkers between [AgF$_x$] moieties.



\begin{figure*}[ht]
\begin{center}
\includegraphics[width=17cm]{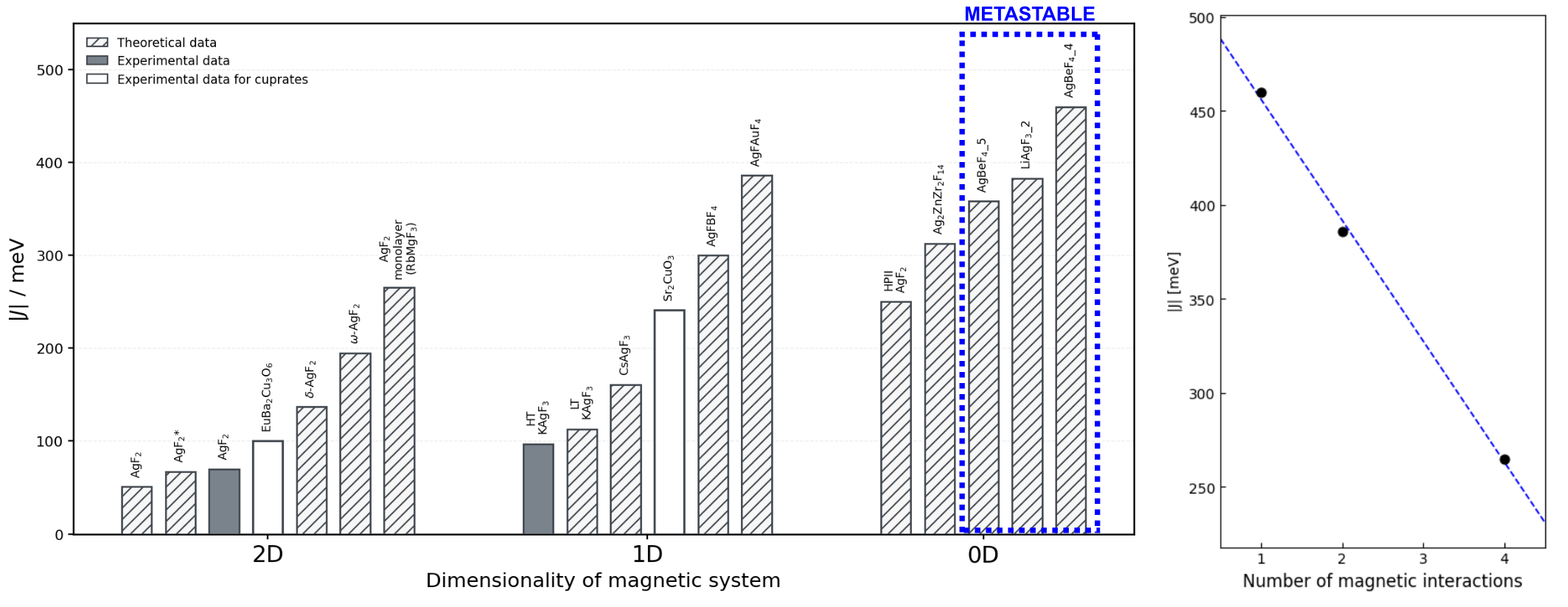}
\caption{LEFT: Comparison of the absolute values of magnetic exchange coupling constants, |J|, for selected Ag(II)-based fluoride systems and representative cuprates, grouped according to the dimensionality of the magnetic lattice. Hatched bars denote theoretical data, filled bars denote experimental data, and open bars represent experimental data for cuprates. The dashed blue box highlights metastable compounds; RIGHT: Dependence of the largest absolute exchange coupling constant |J| on the number of magnetic interactions for Ag-F systems.}
\label{F4}
\end{center}
\end{figure*}

\begin{table}[t]
\centering
\scriptsize
\setlength{\tabcolsep}{3pt}
\renewcommand{\arraystretch}{1.08}
\caption{Comparison of magnetic exchange coupling constants, $|J|$, for selected fluoride and cuprate systems.}
\label{T2}
\resizebox{\columnwidth}{!}{%
\begin{tabular}{@{}lcll@{}}
\toprule
\textbf{System} & \textbf{|J| [meV]} & \textbf{Ref.} & \textbf{Comment} \\
\midrule
AgF$_2$                         & 51  & \cite{63} & DFT+$U$ \\
AgF$_2$$^{*}$                   & 67  & \cite{64} & SCAN \\
AgF$_2$                         & 70  & \cite{65} & Exp. \\
EuBa$_2$Cu$_3$O$_6$             & 100 & \cite{66} & Exp. \\
$\delta$-AgF$_2$                & 137 & \cite{46} & DFT+$U$ \\
$\omega$-AgF$_2$                & 195 & \cite{46} & DFT+$U$ \\
{[}AgF$_2${]} monolayer         & 265 & \cite{67} & DFT+$U$ \\
HT-KAgF$_3$                        & 97  & \cite{68} & Exp. \\
LT-KAgF$_3$                        & 113 & \cite{11} & HSE06 \\
CsAgF$_3$                       & 161 & \cite{11} & HSE06 \\
Sr$_2$CuO$_3$                   & 241 & \cite{69} & Exp. \\
AgFBF$_4$                       & 300 & \cite{12} & HSE06 \\
AgFAuF$_4$                      & 386 & \cite{70} & HSE06 \\
Ag$_2$ZnZr$_2$F$_{14}$          & 313 & \cite{71} & DFT+$U$ \\
AgF$_2$-HPII                    & 250 & \cite{71} & DFT+$U$ \\
LiAgF$_3$\_2                    & 383 & \cite{54} & DFT+$U$ \\
AgBeF$_4$\_5                    & 368 & This work & DFT+$U$ \\
AgBeF$_4$\_4                    & 460 & This work & DFT+$U$ \\
\bottomrule
\end{tabular}%
}
\end{table}

\subsection*{Dynamic stability}
While all polymorphic forms of \ce{AgBeF4} discussed above make perfect sense from the point of view of their crystal structures, connectivity and chemical bonding patterns, it is of interest whether all of them constitute true local minima on the complex multidimensional potential energy surface. While considering this important aspect one should bear in mind that these structures were not proposed based on crystal structure analogies or a wild guess but rather they originate from the unrestricted structure screening which employs evolutionary algorithms. Thus, in principle, all these structures lack any symmetry elements (\textit{P}1) during structure optimizations. Symmetry of these cells is recognized only at a later stage, and the symmetrized cells are reoptimized, which does not lead to any appreciable energy decrease. Therefore, by definition, the theoretical calculations probe an unrestricted potential energy surface and if the algorithm decides to terminate calculation based on the energy criterion, the final structure should always correspond to a local minimum.
However, in order to fully confirm that we have conducted phonon dispersion calculations for all polymorphs discussed above. It turns out that - aside from tiny numerical artifacts - all structures are dynamically stable (cf.  Electronic Supplement, section \textit{S5}) and they constitute genuine local minima. This renders them as viable synthetic targets in experiments. 
The calculated phonon spectra are a fingerprint for each polymorphic structure and thus they could help in identification of the obtained phase if prepared in the future.

\subsection*{Electronic band structure}
All fluorides of silver(II) known so far are broad band gap insulators \cite{73, 74, 75}. It could thus be expected that also the \ce{AgBeF4} polymorphs studied here will belong to the same family of systems. Thus, we have performed the electronic band structure calculations for all polymorphs studied here (cf. Electronic Supplement, section \textit{S4}). In all cases the conduction band corresponds to the Ag-predominated Upper Hubbard Band (UHB), as typical for Ag(II) fluorides. The valence band is composed of a mixture of F(p) and Ag(d) states, with a small excess of the former. This renders \ce{AgBeF4} a charge-transfer insulator within the Zaanen-Sawatzky-Allen diagram, albeit with substantial covalence of the Ag-F bonding.
The computed band gaps vary between 1.18 eV for AgBeF$_4$\_5 (indirect gap) and 1.73 eV for AgBeF$_4$\_2 (indirect gap); however, since DFT+U calculations tend to underestimate the fundamental band gap, these numbers should be treated rather as lower estimates. Interestingly, these values are smaller than the value computed for \ce{AgF2} using the same methodology (2.18 eV \cite{64}), which - in light of the Maximum Hardness Principle - could indicate their metastable character. We will return to this anon. The overall width of the Upper Hubbard Band is the largest for AgBeF$_4$\_5 (0.65 eV) and the smallest for AgBeF$_4$\_2 (0.12 eV); this is consistent with the dimensionality of these systems (the former hosts an infinite chain with short Ag-F bonds, the latter features isolated \ce{AgF4} squares which intreract weakly with one another).

\subsection*{Magnetic ordering and superexchange constants}

For each structure, ferromagnetic (FM) and antiferromagnetic (AFM) spin configurations were tested, and it turned out that the magnetic ground state for every structure proposed in this work is antiferromagnetic. In the next step, to characterise the strength of AFM interactions between silver(II) atoms, the superexchange constant (J) was calculated for each structure. It turned out that the strongest antiferromagnetic interactions are present in AgBeF$_4$\_4 (J$\approx$ --460meV) and in AgBeF$_4$\_5 (J$\approx$ --359 meV). These values are immense, and the first of them surpasses that of --300 meV computed for \ce{AgFBF4} \cite{58} as well as that of --383 meV, suggested recently for one of \ce{LiAgF3} polymorphs \cite{54}. In view of that, it is needed to link the magnetic properties of AgBeF$_4$\_4 and AgBeF$_4$\_5 to their crystal structures (Figure 2, for other polymorphs cf. ESI). As far as AgBeF$_4$\_4 is concerned, its calculated J value within the [Ag$_2$F$_7$]$^{3-}$ dimer surpasses those for structurally related \ce{Ag2ZnZr2F14} (--313 meV) and HP2-form of \ce{AgF2} (--250 meV) \cite{62, 71}. Considering the Goodenough-Kanamori rules \cite{60, 61}, this stems from unusually short Ag-F distance of 2.003\AA~~for AgBeF$_4$\_4  as well as from a nearly linear \ce{Ag(II)\bond{-}F\bond{-}Ag(II)} bridge (178.3$^\circ$). On the other hand, the huge J value for AgBeF$_4$\_5 must to be linked to the presence of unprecedented straight infinite chains, [AgF$_{2/2+2/1}$]$^{-}$, in this structure, as discussed above. Here, the silver(II)--fluorine bonds are also very short. In Figure \ref{F3} the AFM ground state models of the AgBeF$_4$\_4 and AgBeF$_4$\_5 structures are shown with the J interaction marked and the density of spins within the dimers or infinite chains, respectively. The patterns of the spin density are similar to that for related \ce{Sr2CuO3} (with J approx. --240 meV), which is the most strongly coupled system so far confirmed experimentally. However, the superexchange interaction is greater in silver(II) fluoride phase, largely due to marked covalence of the Ag-F bridge \cite{59}. Indeed, the formally closed-shell fluoride anion is immensely spin-polarized, with the excess of the alpha spin density on one side, and the excess of the beta one on the other. While our calculations have utilized a generally accepted U value of 5 eV \cite{47,70, 77}, the robustness of our results was additionally checked via calculations with the Mott-Hubbard U parameter serving as a variable within the 4-6 eV limits (cf. ESI).

Let us now describe the magnetic properties of the less strongly coupled phases. The AgBeF$_4$\_1 structure exhibits in its structure tilted $[AgF_{2/2+2/1}]^{2-}$ chains with the interactions between Ag(II)--F being between short bonds and the \ce{Ag(II)\bond{-}F\bond{-}Ag(II)} angle being 135.9$^\circ$, which leads to the J$\approx$ --129meV, which is comparable with J for other silver(II) containing materials with a similar geometry of the kinked chains \cite{46, 47, 48}.

For AgBeF$_4$\_2 and AgBeF$_4$\_3, superexchange constants are much smaller (J$\approx$ --13meV and --10meV, respectively). The small J value for two polymorphs is due to the interactions between two silver(II) cations being between one long and one short Ag(II)--F bond and at the same time, the isolated squares being oriented almost perpendicular to each other with \ce{Ag(II)\bond{-}F\bond{-}Ag(II)} being 101.2$^\circ$ and 102.2$^\circ$ for AgBeF$_4$\_2 and AgBeF$_4$\_3. All these observations agree well with what is expected based on the~~Goodenough-Kanamori rules \cite{60, 61}.

\subsection*{Magnetic properties of the Ag-Be-F systems as compared to other strongly coupled antiferromagnets}
The discussed two huge J values computed for AgBeF$_4$\_4 and AgBeF$_4$\_5 turn out to be record breaking among all known chemical compounds (Table \ref{T2}, Figure \ref{F4}). In particular, these values exceed those computed previously or measured experimentally for all other Ag(II) fluoride systems and copper(II) oxides. Such outstanding strong superexchange coupling renders the AgBeF4 polymorphs as extremely desirable synthetic targets \cite{11, 26, 39, 46, 47, 48, 54}. This is especially important since magnetically-driven room-T$_C$ \\superconductivity is expected to emerge in metals with antiferromagnetic parent systems showing ---J of approx. 400-700 meV \cite{76}. Our results indicate that the bottom threshold of 400 meV could indeed be surpassed in metastable (as it is discussed in the next section) Be-based stoichiometries studied here, while the upper limit of this property is currently unknown. Interestingly, there also seems to exist a relationship between the maximum attainable |J| value and the magnetic dimensionality, i.e., the smaller the number of magnetically coupled neighbours, the larger the strength of the superexchange interaction. This is reminiscent of the "trans effect" observed for the strength of the chemical bonding, as well as for the primary and secondary bonding in Jahn-Teller active species. The J vs. number of neighbours trend is worthwhile to be verified in experiments.

\begin{figure}[h]
\begin{center}
\includegraphics[width=8.6cm]{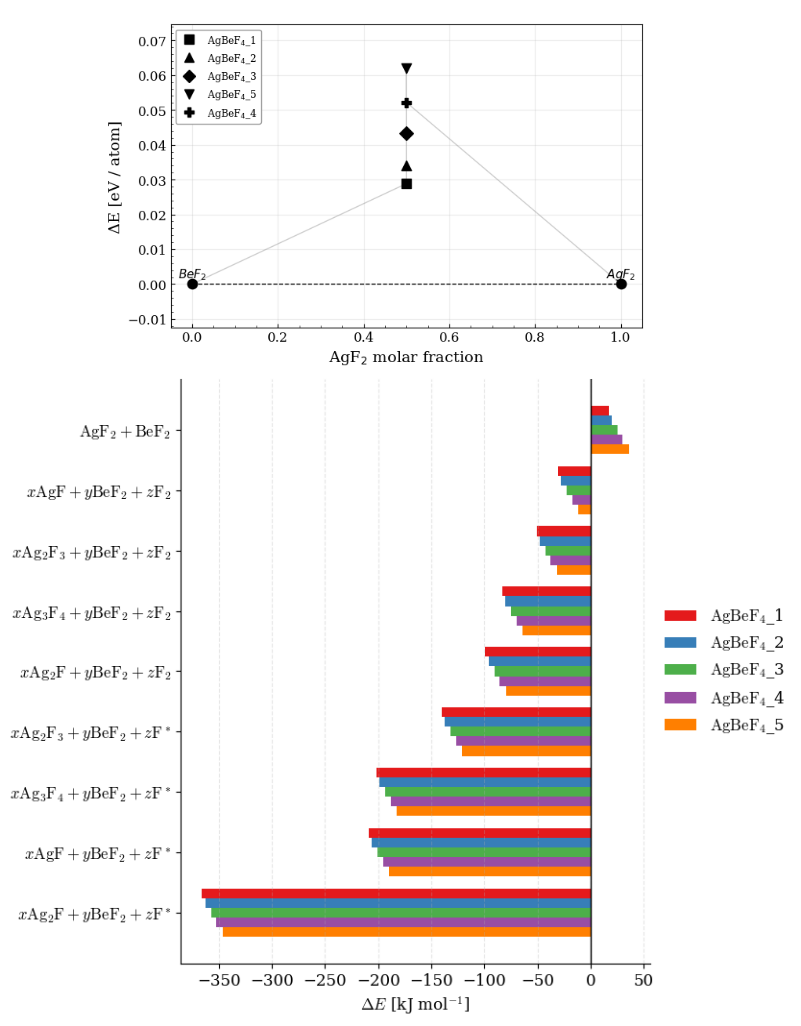}
\caption{TOP: Convexhull for the proposed AgBeF$_4$ polymorphs which shows the relative stability of the considered structures with respect to AgF$_2$ and BeF$_2$. BOTTOM: Calculated formation energies of these structures from selected precursors in reactions involving fluorine gas (F$_2$) and fluorine radicals (F$^\ast$).}
\label{F2}
\end{center}
\end{figure}

\subsection*{Energetic stability and possible synthetic pathways}
To assess whether the calculated structures are thermodynamically stable in regards to the binary fluorides: AgF$_2$ and BeF$_2$, a convex hull was constructed (Figure \ref{F2}). As shown in Figure \ref{F2}, all polymorphs lie slightly above the convex hull with the energy per atom ranging from approx. +30meV to +60meV. This indicates that these structures are metastable and likely could not be obtained via direct synthesis from the binary fluorides under standard conditions. However, such small energy differences above the hull often suggest that these phases could be trapped as local minima on the potential energy surface if reactions proceed from higher-energy substrates. Thus, we explored  alternative synthetic routes and we calculated the reaction enthalpy for various sets of precursors. These encompassed silver containing compounds such as: Ag$_2$F, AgF, Ag$_2$F$_3$ and Ag$_3$F$_4$ with BeF$_2$ and both F$_2$ (fluorine gas) and fluorine radical (F$^*$) (AgF$_3$ is unstable so it was excluded \cite{49}). Many exotic and metastable phases were prepared using fluorine radicals in anhydrous HF in the past. As presented in Figure \ref{F2} bottom, the reactions involving the fluorine radical are much more exothermic than reactions with F$_2$, as expected. Indeed, both reactive agents are commonly used in the synthesis of metastable compounds \cite{50, 51, 52}. The reaction involving Ag$_2$F, BeF$_2$ and the fluorine radical is the most exothermic (approx. -370 to -350 kJ/mol depending on the targetted polymorph), suggesting a strong thermodynamic driving force. This high energy gain could potentially overcome the activation barriers required to stabilize the AgBeF$_4$ framework, making preparation realistic. After all, they key craftsmanship of chemists consist of providing reaction conditions needed to stabilize plethora of metastable phases prepared by mankind and despite the existence of many thermodynamic sinks.

We also notice that the fact that polymorphs of \ce{AgBeF4} with different forms of the anionic fluoroberyllate sublattices (isolated anions, dimers, infinite chains) have comparable energies testifies to comparative Lewis acidities of \ce{AgF2} and \ce{BeF2}. Indeed, with group 1 metal fluorides \ce{AgF2} forms salts featuring the polymeric AgF$_3$$^-$ anion, hence \ce{AgF2} serves here as a Lewis acid. On the other hand, with a group 3 fluoride, \ce{BF3}, it forms the \ce{AgFBF4} and \ce{Ag(BF4)4} salts (the latter in the aHF solutions \cite{58}), hence \ce{AgF2} serves here as a Lewis base. The case of \ce{BeF2}, a group 2 metal fluoride, is intermediate between these two extremes, and the Ag(II) and Be(II) centers compete for the fluoride anions. This fact allows to explain why the Be-F bond lengths in the phases studied here are somewhat longer by up to 0.05 \AA~~ than those found in fluoroberylates of more electropositive elements.

\section*{Conclusions}
\label{conclusion}

Here we presented the first systematic computational investigation of the  landscape of structural polymorphism, dynamic and energetic stability, and magnetic and electronic properties of AgBeF$_4$ crystalline phase. Our results demonstrate that the combination of the small, rigid BeF$_4$ tetrahedra with the electronically active $Ag^{2+}$ ion leads to structural motifs distinct from those observed in heavier alkaline-earth XBeF$_4$ analogs. The AgBe$F_4$\_4 polymorph (J$\approx$ --460\text{ meV}) featuring AgF$_2$F$_7$ dimers confirms that beryllium fluorides can serve as excellent scaffolds for stabilizing low-dimensional magnetic systems with record-high~~superexchange constants. On the other hand, AgBe$F_4$\_5 (J$\approx$ --368\text{ meV}) hosts perfectly straight chains composed of equiplanar [\ce{AgF4}] squares which are unprecedented in the entire Ag(II) fluoride chemistry. Thus, the current study underscores the utility of global structure prediction in uncovering hidden phases in unexplored fluoride systems and provides a theoretical roadmap for the synthesis of new silver(II)-based magnetic materials. 
\\While the AgBeF$_4$ phases are metastable relative to AgF$_2$ and BeF$_2$, the large exothermic driving force calculated for reactions of their synthesis involving Ag$_2$F and fluorine radicals, as well as their dynamic (phonon) stability points toward a feasible experimental path for their realization. The true limits of the strength of the magnetic superexchange in these materials is still unknown and will require the experimental verification. However, such studies are worthwhile as the metastable small-band gap phases like those studied here may provide particularly immense coupling constants.

\section*{Experimental}
\label{experimental}

The structural screening was carried out using the XtalOpt r.13.2 package \cite{1, 2, 16}, which incorporates self-learning algorithms. Initial structures were generated randomly, for unit cells containing 2, 4, 6 and 8 formula units. In each search, a total of 650 structures were produced, including 100 initial seed configurations.

All calculations were performed with VASP 5.4.4 (Vienna Ab initio Simulation Package) \cite{3,4,5}. The interaction between valence electrons and ionic cores was described using the projector augmented--wave (PAW) method \cite{6, 7}, with pseudopotentials constructed within the PBE formalism. Exchange correlation effects were treated using the generalized gradient approximation (GGA) with the PBEsol functional \cite{8}. A plane-wave cut-off energy of 600eV was applied, and the self-consistent-field convergence threshold was set to $10^{-7}$ eV.

Fifteen lowest-enthalpy unique structures identified in each search were selected for further optimization. These structures were examined within different ferromagnetic and antiferromagnetic configurations using the DFT+U approach, as implemented in the Dudarev scheme \cite{9}, to properly account for localized d electrons. The on-site Coulomb and exchange parameters were set to U = 5.0eV and J = 1.0eV for the Ag d orbitals. The inclusion of the Hubbard $U$ correction $(DFT+U)$ is crucial for an accurate description of the strongly correlated $4d$ electrons in Ag(II) \cite{10, 11, 12}. The optimized structures were symmetrized using FINDSYM v.7.1.7 \cite{13, 14}. After this step, the five lowest energy polymorphs were selected for further studies.

The direct and indirect band gaps, as well as the density of states (DOS), electronic band structure, and phonon dispersion, were calculated using the SUMO software \cite{15}.

\section*{Acknowledgements}

This work was supported by the Polish National Science Center (project  2024/53/B/ST5/00631). Computations were performed at the ICM (University of Warsaw, GA83-34) and Wroc{\l}aw Centre for Networking and Supercomputing (grant no. 484). 

\section*{Conflict of Interest}

The authors declare no conflict of interest.

\begin{shaded}
\noindent\textsf{\textbf{Keywords:} \keywords} 
\end{shaded}


\setlength{\bibsep}{0.0cm}
\bibliographystyle{Wiley-chemistry}
\bibliography{example_refs}

\clearpage


\section*{Entry for the Table of Contents}



\noindent\rule{11cm}{2pt}
\begin{minipage}{5.5cm}
\includegraphics[width=5.5cm]{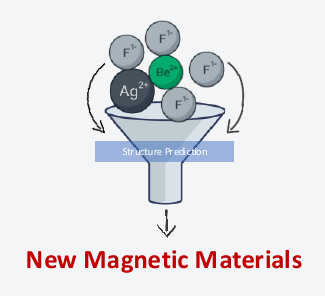} 
\end{minipage}
\begin{minipage}{5.5cm}
\large\textsf{A computational search reveals five metastable AgBeF$_4$ polymorphs. The $P\bar{1}$ AgBeF$_4$\_5 phase features unprecedented $[AgF_{2/2+2/1}]^{2-}$ chains with a record strong antiferromagnetic superexchange coupling (J= --460 meV) between Ag(II) centers.}
\end{minipage}
\noindent\rule{11cm}{2pt}

\vspace{2cm}




\includepdf[pages=-]{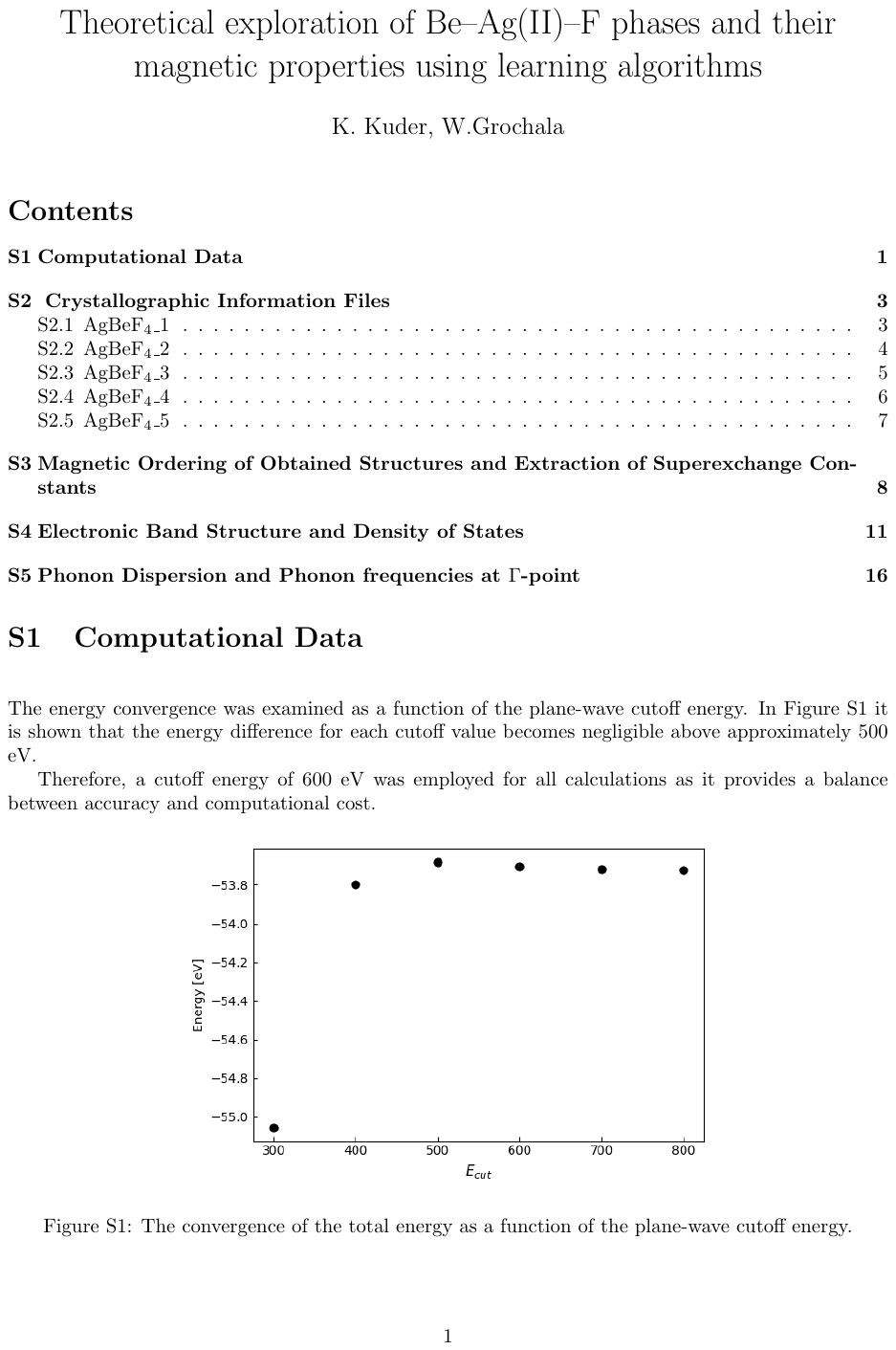}

\end{document}